\documentclass{aa}  
\usepackage{graphicx}
\usepackage{tikz}
\usepackage{txfonts}
\usepackage{hyperref}

\begin{document}

   \title{Contact binary asteroid (153201) 2000 WO$_{107}$:\\rotation, shape model, and density}

   \author{Yurij Krugly\inst{1,2,3}\thanks{\href{e-mail: }{krugly@astron.kharkov.ua}}
   \and
   Oleksiy Golubov\inst{2,3}
   \and
   Ihor Kyrylenko\inst{2}
   \and
   Veronika Lipatova\inst{4,2}
   \and
   Irina Belskaya\inst{2}
   \and
   Vasilij Shevchenko\inst{2}
   \and
   Ivan Slyusarev\inst{2}
   \and
   Raguli Inasaridze\inst{5,6}
   \and
   Shuhrat Ehgamberdiev\inst{7,8}
   \and
   Oleksandra Ivanova\inst{9,10}
   \and
   Marek Hus\'{a}rik\inst{9}
   \and
   Sergey Karpov\inst{11} 
   \and
  Daniel Hestroffer\inst{1}
   } 

   \institute{LTE, Observatoire de Paris, Universit\'{e} PSL, Sorbonne Universit\'{e},
Universit\'{e} de Lille, LNE, CNRS, 75014 Paris, France
    \and
    Institute of Astronomy,
V.N. Karazin National University, 35 Sumska Str., Kharkiv, 61022, Ukraine 
    \and
    Astronomical Observatory Institute, Faculty of Physics, Adam Mickiewicz University, Poznań, PL-60-286, Poland 
    \and 
    Institut f\"ur Theoretische Astrophysik, Zentrum f\"ur Astronomie, Universit\"at Heidelberg, D-69120 Heidelberg, Germany
    \and
    E. Kharadze Georgian National Astrophysical Observatory, Abastumani, Georgia
    \and
    Samtskhe-Javakheti State University, Akhaltsikhe, Georgia
    \and
    Ulugh Beg Astronomical Institute of the Uzbek Academy of Sciences, Tashkent, Uzbekistan
    \and National University of Uzbekistan, Tashkent, Uzbekistan
    \and
    Astronomical Institute of the Slovak Academy of Sciences, SK-059 60 Tatransk\'{a} Lomnica, the Slovak Republic
    \and
    Main Astronomical Observatory of the National Academy of Sciences of Ukraine, Kyiv, Ukraine
    \and
    Institute of Physics of the Czech Academy of Sciences, CZ-182 21 Prague 8, Czech Republic
    }

  \abstract
   {Near-Earth asteroid (153201) 2000 WO$_{107}$ exhibits spectral properties and albedo consistent with the taxonomic type M, which implies the possibility of high metal abundance and larger density.}
   {We combine different methods to investigate the asteroid's rotation, determine its shape, and use it to estimate its density.}
   {We carried out photometric observations of the asteroid during the 2020 apparition. Then we created a program able to simulate the lightcurves, and used it within a Markov chain Monte Carlo (MCMC) algorithm to reconstruct the asteroid shape model from the observational data. The Goldstone radar observations of the asteroid were used as an additional constraint on the asteroid model in the MCMC algorithm. The estimated shape and rotation rate of the contact binary were used to compute its density.}
   {The photometric observations of (153201) 2000 WO$_{107}$ obtained at a wide range of the phase angles from 5 to 68 degrees in the time interval November 28 -- December 8, 2020, show lightcurves typical for contact binary asteroids, which agrees with the results of the radar data. The lightcurves have a maximum amplitude of up to 1.24 mag. 
   The best-fit modelled shape of the asteroid is composed of two ellipsoidal lobes with the axes $0.68\times 0.38 \times 0.36$ km and $0.44 \times 0.42 \times 0.16$ km. Its sidereal rotation period is determined to be $5.017\pm 0.002$ hr. The most probable solution for the angular velocity vector of the asteroid points at the ecliptic coordinates $\lambda=96^\circ \pm 8^\circ$ and $\beta=-78^\circ \pm 1^\circ$, whereas another less probable solution around $\lambda=286^\circ \pm 11 ^\circ$, $\beta=-76^\circ \pm 2 ^\circ$ cannot be disregarded. 
   The estimated density of the asteroid $\rho=4.80^{+0.34}_{-0.63}$ g/cm$^3$ is consistent with its possible metallic composition. From the orbital simulation of this potentially hazardous asteroid, we find that its integral probability of colliding with the Earth in the next 10,000 years is $7\cdot 10^{-5}.$}
   {}


   \maketitle

\section{Introduction}
Near-Earth potentially hazardous asteroid (153201) 2000 WO$_{107}$ (which we call WO$_{107}$ for brevity) was discovered by LINEAR at Socorro in December 2000. Its orbit has a significant eccentricity $e = 0.781$ and crosses the orbits of all terrestrial planets. The asteroid intersects the orbit of Mercury and approaches the Sun at a small perihelion distance of $q = 0.20$  AU and an inclination $i = 7.77$ deg. WO$_{107}$ is a potentially hazardous asteroid, with the minimum orbit intersection distance with the Earth MOID = 0.00289 AU\footnote{Minor Planet Center, List Of Potentially Hazardous Minor Planets, \url{https://minorplanetcenter.net/iau/mpc.html}}. It passes closely near the Earth every 20 years.

The asteroid was classified as an X-type from spectral observations \citep{binzel2004}. According to NEOWISE data its diameter is $510 \pm 83$ m and albedo $p_\mathrm{V} = 0.129 \pm 0.058$ \citep{mainzer2019}. The ambiguous spectral X-type encompasses E-, P- and M-types, which are distinguished from each other by albedo. Given the known moderate albedo, the asteroid is most consistent with the M-type. In the population of near-Earth asteroids, the abundance of M-type asteroids is smaller 1--3\% \citep{binzel2019}. 

In the 2020 opposition, WO$_{107}$ passed at the minimum distance of 0.029 AU from the Earth, with its visible magnitude reaching 13.2 mag (MPC IAU data). At its close approach, the asteroid was observed under rapidly changing observation geometry, as its phase angle changed over several days from more than 100 degrees to 3 degrees. Such observation conditions potentially make it possible to obtain lightcurves at different geometry and create a model of its body shape. We initiated photometric observations of WO$_{107}$ to obtain its rotation parameters and estimate its shape and bulk density.

During the close approach, the asteroid was also the target of radar observations, which obtained its images with a resolution of up to 19 m/pixel \citep{benner}. 

\cite{franco2021} published the high amplitude composite lightcurve of WO$_{107}$ which was observed during 2 nights on Nov 29 and Dec 14, 2020. \cite{warner2021} observed WO$_{107}$ on Dec 6-12 and also obtained the high amplitude lightcurves with particular minima: one sharply deep and another shallow with a sloped plateau.  

In Section \ref{sec:observations} we present the results of our photometric observations of WO$_{107}$.
In Section \ref{sec:shape} we supplement the optical photometry with the radar images and use these data to model the asteroid's shape.
In Section \ref{sec:density} the shape is used to estimate the asteroid density.
In Section \ref{sec:origin} the origin and orbital evolution of WO$_{107}$ is considered.

\section{Photometric Observations}
\label{sec:observations}

Our photometric observations were carried out within a coordinated program at five observatories with the 70-cm telescope at the Abastumani Observatory, the 1-m telescope at the Simeiz Observatory, the 61-cm telescope at the Skalnat\'{e} Pleso Observatory, the 36-cm telescope at Kitab Observatory, and the 25-cm BART (now called FRAM-ORM) telescope at the Roque de los Muchachos Observatory. The observations were done using CCD cameras through mainly $R$ and additionally $BVI$ filters of the Johnson-Cousins system, and without a filter in Kitab and Simeiz. Table \ref{tab:observatories} contains parameters of the telescopes and cameras used.

\begin{table*}
\centering 
\caption{The telescopes and cameras used for observations of the asteroid (153201) 2000 WO$_{107}$.} 
\begin{tabular}{|l|l|c|c|c|c|}
\hline
MPC code and Observatory & Telescope & FOV & Camera & Resolution  & Binning \\
 & & [arcmin] & & [px] & [px]\\
\hline
(119) Abastumani, Georgia & 70-cm Maksutov meniscus & $45\times 45$ & FLI PL4240 & 2k $\times$ 2k & $1 \times 1$ \\

(094) Simeiz, Ukraine	& 1-m Ritchey--Chretien	& 10 $ \times $ 10 &	FLI PL09000 &  3k $\times$ 3k & $3\times 3$ \\
 
(186) Kitab, Uzbekistan	& 36-cm Cassegrain	& 44 $ \times $ 44 &	FLI PL09000 & 3k $\times$ 3k & $2 \times 2$ \\
 
(056) Skalnat\'{e} Pleso, Slovakia	& 61-cm Newtonian	& 19 $ \times $ 13 & SBIG ST-10XME & 2k $\times$ 1.5k & $2 \times 2$ \\

(J04) Roque de los Muchachos, & 25-cm BART & $25\times 25$ & CCD 47-10 & 1k $\times$ 1k & $1 \times 1$\\
\,\,\,\,\,\,\,\,\,\,\,\,\,\,Tenerife, Spain & & & & & \\
\hline
\end{tabular}
\label{tab:observatories}
\end{table*}

The observations aimed to obtain lightcurves of WO$_{107}$ over a wide range of aspect and phase angles for about two weeks during its close approach in late November -- early December 2020. The asteroid came to the Earth from the side of the Sun, and on November 29 it crossed the Earth's orbit, making its closest passage. Its illumination geometry with respect to the observers was very rapidly changing during the passage. Over several nights, the solar phase angle of the asteroid changed from more than 90 degrees to the minimum angle of 3 degrees on December 2. On the following nights in December, the asteroid rapidly moved away from the Earth, losing brightness and slowly changing the aspect and the solar phase angle. The proximity of the asteroid to the bright full moon obstructed the possibility of observing it on two nights, November 30 and December 1.

All observations have been carried out by telescopes with sidereal tracking, except the 36-cm telescope at Kitab, which utilized asteroid tracking. In sidereal mode, the telescopes lag behind the fast asteroid, so they have to be moved to a part of the field of view (FOV) from time to time to get some overlap with the previous field and to combine them using mutual comparison stars \citep{krugly2004}. The asteroid was sufficiently bright to obtain good-quality photometry in several filters. The primary reduction of the observed images of WO$_{107}$ includes the removal of master dark and normalization by master flat-fields. The master flat-fields were obtained by a median combination of the twilight images for each used filter and for the unfiltered mode. If the sky background on raw observed images had a slope due to the moonlight, it was corrected by a linear gradient filter. Overall, our observations lasted from November 28 to December 8, as it is presented in Table 2.

The measurements of the asteroid's magnitude were done with aperture photometry using two software programs. Firstly, we used the AstPhot package \citep{mottola1995} to perform differential photometry of the asteroid relative to the nearest comparison stars (more details in \citealt{krugly2002}).
The second one, the MPO Canopus software was used to obtain the "derived" calibrated magnitude \citep{warner2022} with a technique that included selecting comparison stars whose spectra were close to the solar one, that is, close to the reflectance spectrum of the asteroid. The MPO Canopus supposes a possibility to use comparison stars with known magnitudes in the photometric system, which can be different from the BVRI Johnson-Cousins system, but demands checking errors in the transformation between the systems. We have compared the determined magnitudes of the asteroid, which were obtained using up to 5 chosen comparison stars observed on the same images, and were taken directly or transformed to BVRI magnitudes from several photometric catalogs included in the Canopus: APASS, CMC15, and ATLAS. The estimated errors of the calibrated magnitudes of the comparison stars are in the range of 0.02-0.05 mag. 

The size of the aperture is adjusted to partly or fully cover the image of the asteroid and comparison stars. When the asteroid is bright, the aperture size is chosen to cover more than 95\% of the measured object’s brightness. If the asteroid is not sufficiently bright, the efficient aperture radius is selected to maximize the signal-to-noise ratio (SNR), which could be estimated from the growth curves for bright stars on the measured images. The circular aperture works efficiently only for slow-moving asteroids or short exposure times when the asteroid looks like a fixed star. In this case, the same size of the aperture is used for the asteroid and the comparison stars. The elliptical apertures are used for the asteroid's images, elongated in the direction of its motion. The small semi-axes of these apertures should be equal to the radius of the efficient circular aperture of the comparison stars. Images of the asteroid observed at the Kitab Observatory with the 36-cm telescope with the asteroid tracking mode look like a fixed star, and they were measured with a circular aperture. In this case, elliptical apertures were used for measuring the stretched images of the comparison stars. 

\begin{figure}
\centering
\includegraphics[width=0.49\textwidth]{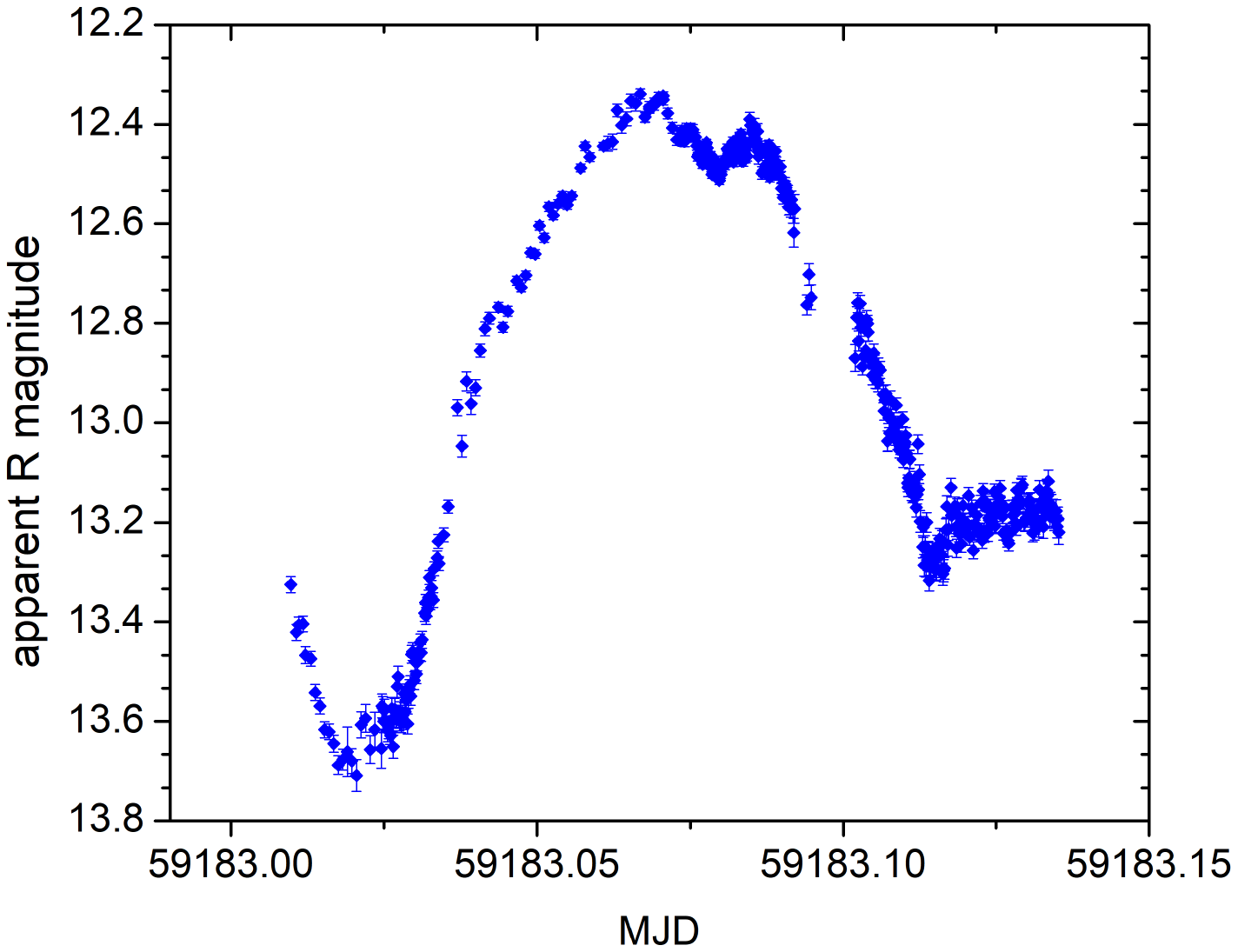}
\caption{The lightcurve of WO$_{107}$ with apparent magnitudes observed at Simeiz Observatory on November 29/30, 2020.}
\label{fig:ligcurve2}
\end{figure}

The observations at the Simeiz Observatory were made with $10 \times 10$ arcmin FOV is inconveniently small for observations of an asteroid moving as fast as 1 arcmin per minute. On November 28 and 29 we observed the asteroid with short 10, 20, or 30 s exposure times and obtained 10 -- 30 images of the asteroid in one FOV, in which we could use the same comparison stars for differential photometry. The aperture photometry of these observations was performed with two methods using different software. The unfiltered observations were measured using the AstPhot software. The obtained short relative lightcurves measured for one FOV were combined using comparison stars that were present in both neighboring fields. The other method was used for measuring observations in the R filter of the Johnson-Cousins system with the MPO Canopus software. We did calibrated photometry in each FOV using up to five comparison stars with solar-like colors, resulting in short calibrated lightcurves. The values of the $R$ magnitudes of the comparison stars were determined by transformation from CMC15 or/and ATLAS catalogs \citep{warner2022}. The $R$-filter lightcurve measured with the Canopus is shown in Figure \ref{fig:ligcurve2}. Up to 40 FOVs were individually measured to construct a lightcurve. Fluctuations from one group to another are related to the accuracy of the calibration. The average deviation is at the level of 0.02-0.05 mag, and the maximal errors are up to 0.1 mag. The most anomalous
``short'' lightcurves were removed. 

The first observations of the asteroid on November 28 began a few hours before the nearest passage to the Earth early on November 29 at high phase angles and with very fast proper motion. These observations were carried out at three observatories: Kitab, Abastumani, and Simeiz (see Table \ref{tab:observations}).

\begin{table*}
\centering
\caption{Observation log and aspect data of asteroid (153201) 2000 WO$_{107}$.}   
\begin{tabular}{|c|c|c|c|c|c|c|c|c|l|}
\hline
UT Date & Duration UT& Filter(s) & $V$ & $\lambda_\mathrm{PAB}$ & $\beta_\mathrm{PAB}$ & $r$ & $\Delta$ & $\alpha$ & Observatory \\
(2020) & [hrs] & & [mag] & [deg] & [deg] & [AU] & [AU] & [deg] & \\
\hline
11 28.9 & 19.92 -- 01.33 & Clear & 14.02 & 102.67 & --2.41 & 0.997 & 0.029 & 68.7 & Kitab \\
11 29.0 & 21.48 -- 23.07 & R & 13.92 & 101.28 & --2.22 & 0.998 & 0.029 & 65.7 & Simeiz \\
11 29.0 & 22.51 -- 01.46 & R & 13.92 & 101.28 & --2.22 & 0.998 & 0.029 & 65.7 & Abastumani \\
11 29.9 & 17.35 -- 22.47 & Clear & 13.32 & 89.01 & --0.61 & 1.009 & 0.030 & 40.1 & Kitab \\
11 29.9 & 20.66 -- 23.51 & Clear & 13.32 & 89.01 & --0.61 & 1.009 & 0.030 & 40.1 & Simeiz \\
11 30.0 & 0.23 -- 03.25 & R & 13.28 & 87.77 & --0.46 & 1.011 & 0.031 & 37.5 & Simeiz \\
11 30.0 & 22.37 -- 02.67 & R & 13.28 & 87.77 & --0.46 & 1.011 & 0.031 & 37.5 & Abastumani \\
12 02.9 & 18.50 -- 0.13 & BVR & 13.72 & 68.62 & 1.75 & 1.046 & 0.060 & 5.0 & Skalnat\'{e} Pleso \\
12 02.9 & 21.66 -- 22.67 & VR & 13.74 & 68.62 & 1.75 & 1.046 & 0.061 & 5.2 & Abastumani \\
12 04.0 & 19.08 -- 06.37 & BVR & 14.44 & 65.99 & 2.06 & 1.060 & 0.076 & 11.1 & Roque de los Muchachos \\
12 04.0 & 22.40 -- 02.25 & BVRI & 14.49 & 66.03 & 2.05 & 1.059 & 0.075 & 11.0 & Simeiz \\
12 05.0 & 23.89 -- 01.91 & VR & 15.03 & 64.51 & 2.24 & 1.071 & 0.089 & 14.9 & Simeiz \\
12 05.1 & 23.42 -- 04.63 & BVR & 15.11 & 64.40 & 2.26 & 1.072 & 0.090 & 15.2 & Roque de los Muchachos \\
12 08.8 & 18.03 -- 20.93 & BVR & 16.47 & 61.91 & 2.67 & 1.114 & 0.143 & 23.7 & Skalnat\'{e} Pleso \\
\hline
\end{tabular}
\label{tab:observations}
\tablefoot{The columns contain: time and interval of observations in UT, filter(s) used, visible magnitude taken from Minor Planet Center, longitude and latitude of phase angle bisector, distances from the asteroid to the Sun and Earth, solar phase angle, observing site}
\end{table*}

\begin{figure}
\centering
\includegraphics[width=0.49\textwidth]{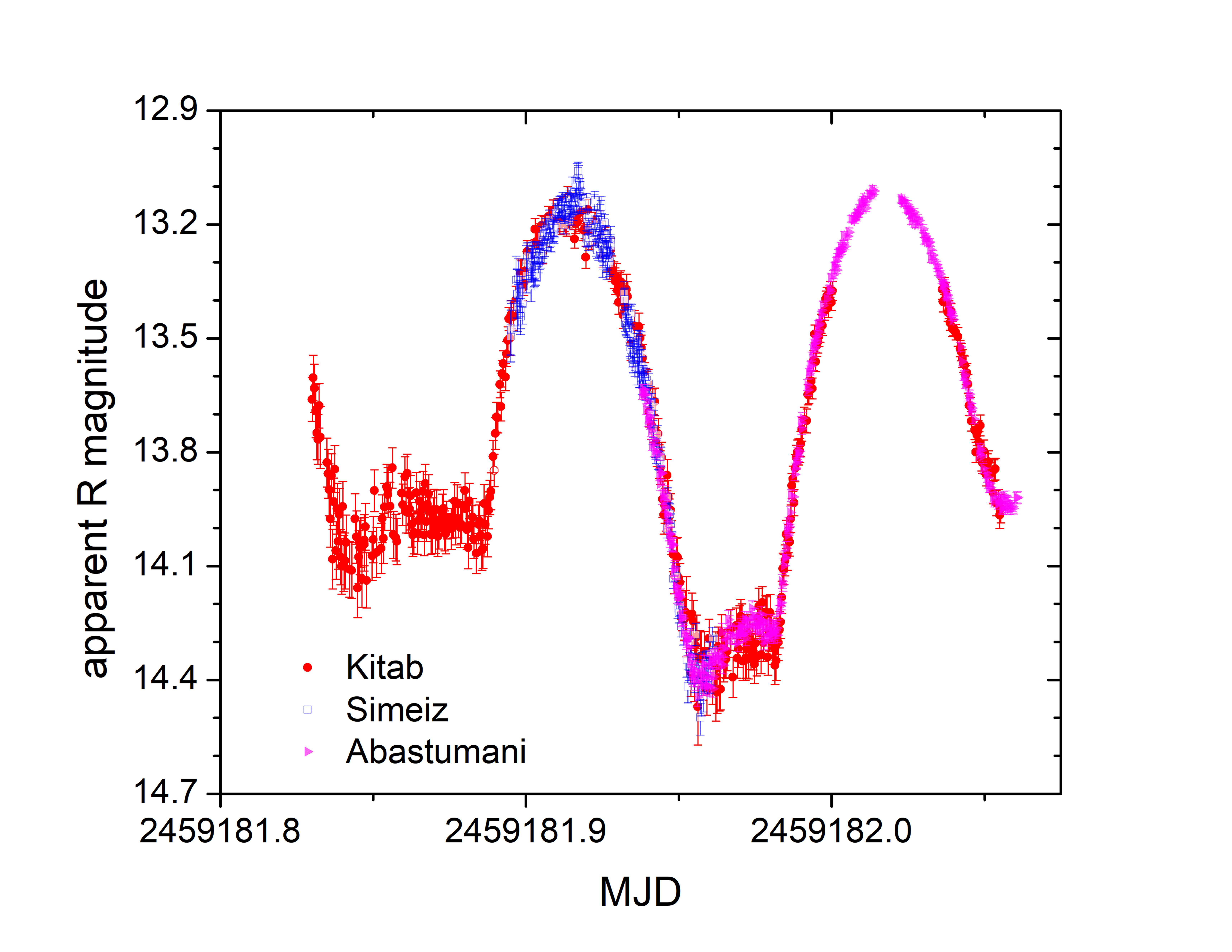}
\caption{The lightcurves obtained from observational data on November 28/29 during the passage of the asteroid at a minimum distance from the Earth.}
\label{fig:ligcurve1}
\end{figure}

The measured lightcurves on November 28 showed a very high amplitude of up to 1.24 mag and made it possible to determine the rotation period of $5.05$ hrs (Figure \ref{fig:ligcurve1}). In part, the high amplitude can be explained by observations at large phase angles of 65-70 deg \citep{zappala1990}. Still, such a high-amplitude lightcurve with deep and flattened minima characterizes the asteroid as a very elongated and possibly binary body. The unusual flat minima are a characteristic manifestation of incomplete illumination of the asteroid at large phase angles. Such illumination conditions can occur for synchronous binaries or very elongated nonconvex bodies such as contact binaries \citep{lacerda2007}, again indicating that the asteroid can be either a binary or a contact binary. 

\begin{figure}
\centering
\includegraphics[width=0.49\textwidth]{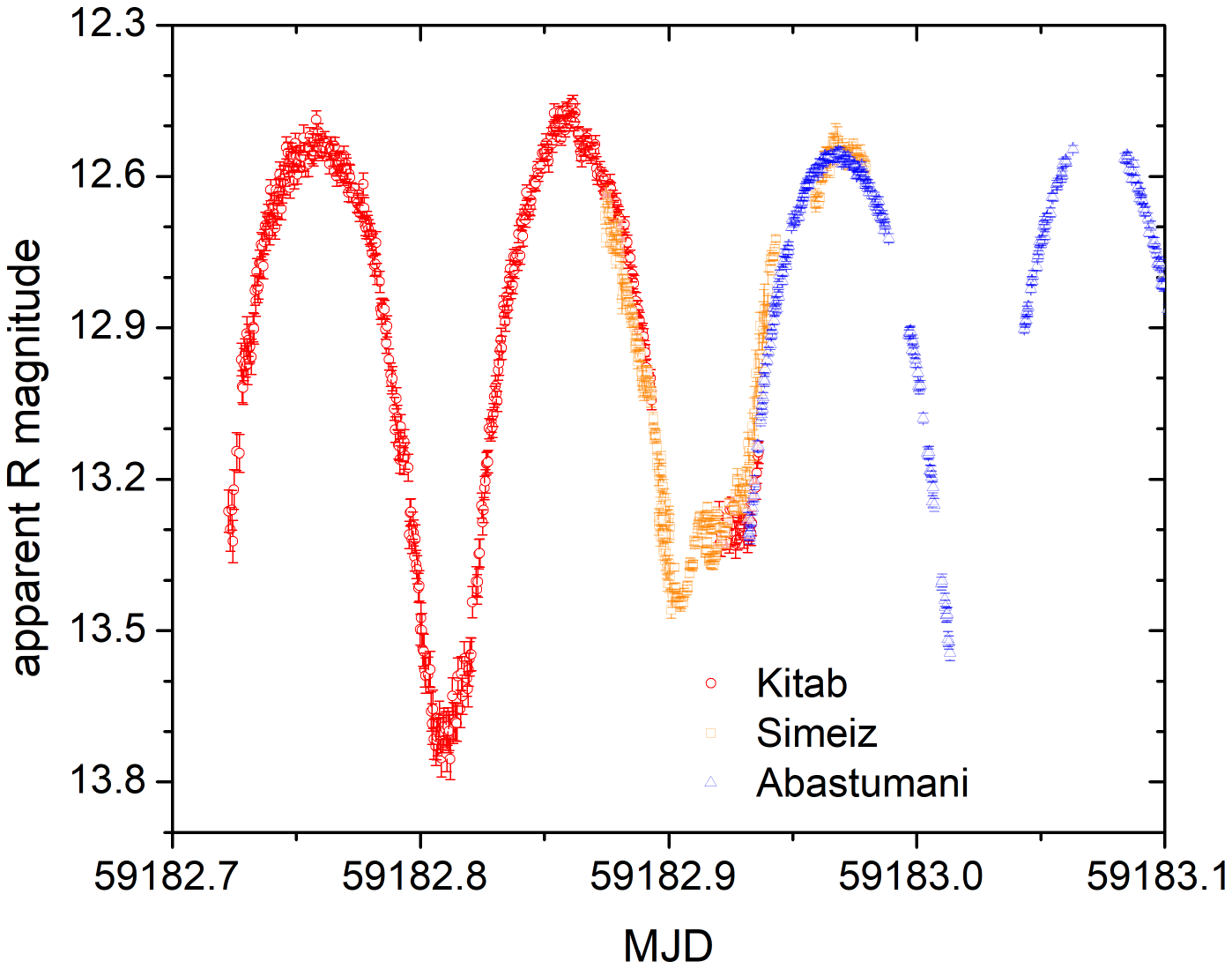}
\caption{The lightcurves of WO$_{107}$ obtained  on November 29/30, 2020.}
\label{fig:composite-Nov29}
\end{figure}

The next night's observations on November 29 the asteroid was observed at 35--40 deg phase angles and confirmed the high amplitude of the lightcurve and unusual shapes of the minima. One of the observed minima of the lightcurve is deep and sharp, and the other one is shallower with a plateau (see Figure \ref{fig:composite-Nov29}). 

Our assumption on the binary nature of WO$_{107}$ was confirmed by radar observations carried out using the 70-m antenna at the Goldstone Observatory at the same opposition. The radar data showed that the asteroid is a contact binary with a larger elongated primary component and a smaller secondary attached to the end of the longest axis of the larger body \citep{benner}. 

\begin{figure}
\centering
\includegraphics[width=0.49\textwidth]{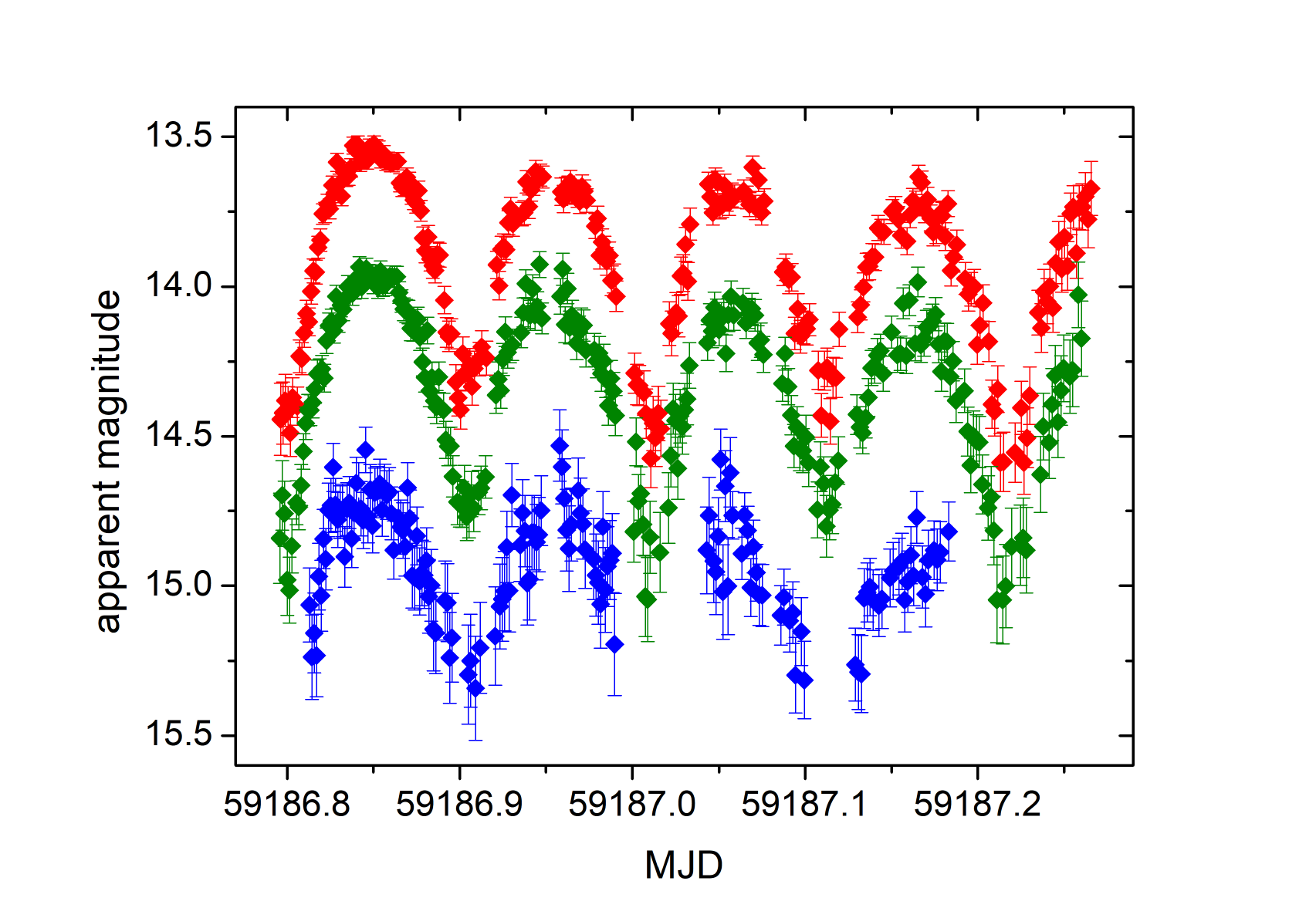}
\caption{The BVR lightcurves of WO$_{107}$ observed with the BART (FRAM-ORM) telescope at the Roque de los Muchachos on December 3/4, 2020. From bottom to top curve: B in blue, V in green, R in red.}
\label{fig:ligcurve3}
\end{figure}

\begin{figure*}
\centering
\includegraphics[width=0.5\textwidth]{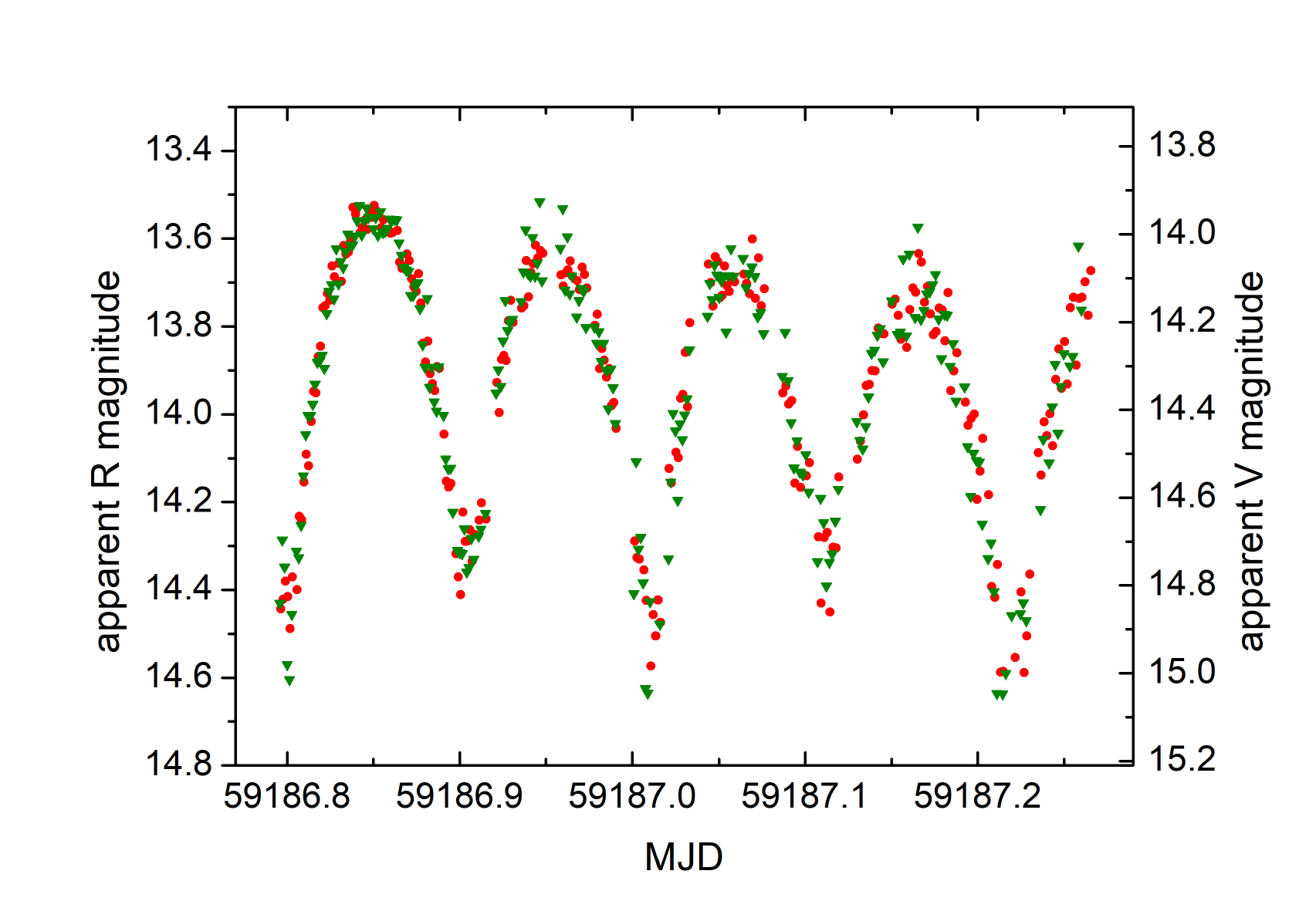}
\includegraphics[width=0.45\textwidth]{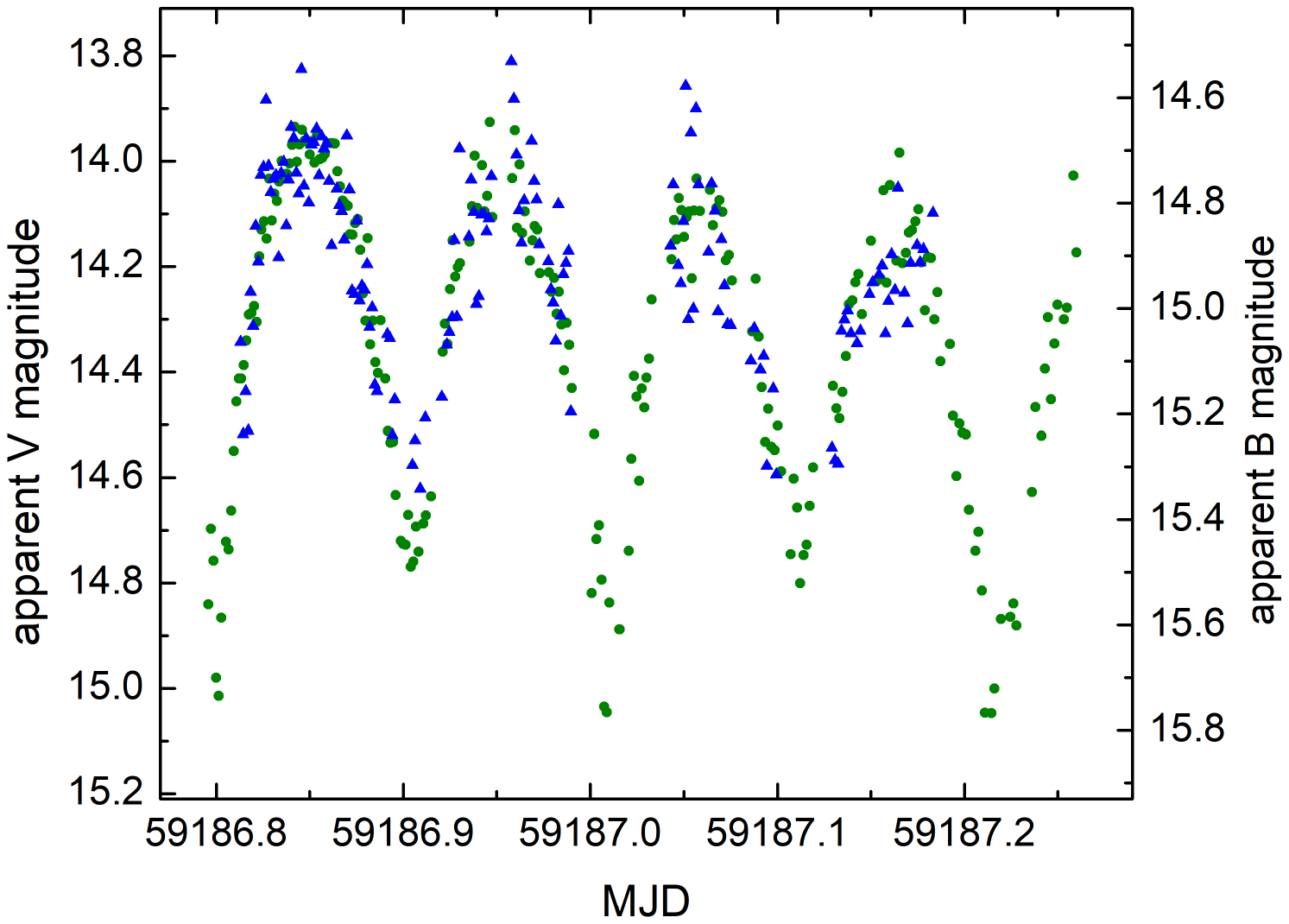}
\caption{The lightcurves of WO$_{107}$ for B and V filters (left) and V and R filters (right), obtained with the BART (FRAM-ORM) telescope at the Roque de los Muchachos on December 3/4, 2020. The lightcurves were shifted along the magnitude axis by the values of the obtained color indices.}
\label{fig:ligcurve4}
\end{figure*}

The observations with the 25-cm BART (FRAM-ORM) telescope at the Roque de los Muchachos Observatory on December 3 were carried out in three filters BVR of the Johnson-Cousins photometric system. These calibrated lightcurve data obtained at the solar phase angle of 11 deg are shown in Figure \ref{fig:ligcurve3}. The data were used to derive the colors of the asteroid: $B$--$V$ = $0.72 \pm 0.11$ mag and $V$--$R$ = $0.41 \pm 0.05$ mag. These colors are consistent with the asteroid having M-type \citep{bowell1978,belskaya2020}. Combinations of lightcurves obtained in different filters are shown in Figure \ref{fig:ligcurve4}.

\begin{figure}
\centering
\includegraphics[width=0.49\textwidth]{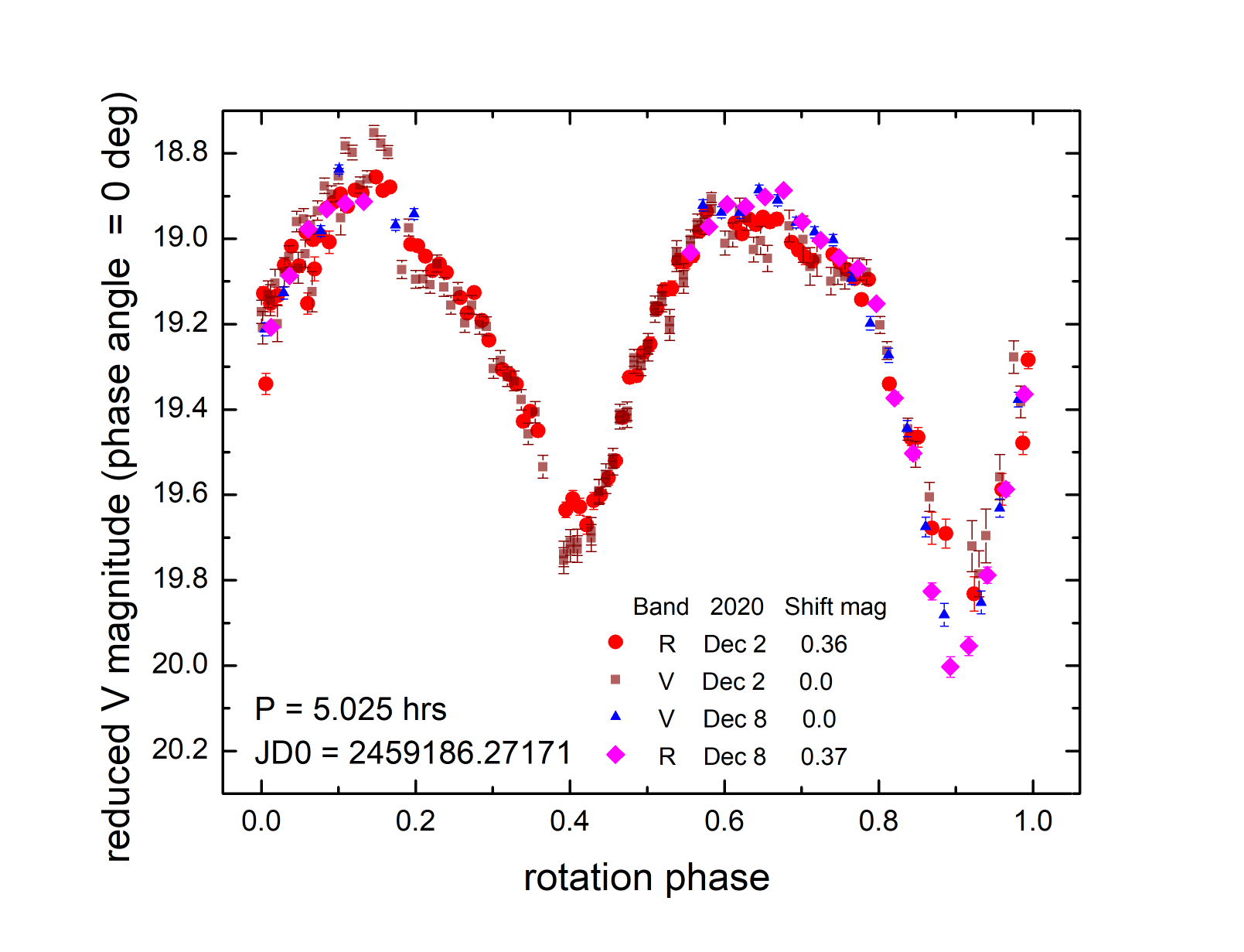}
\caption{The composite lightcurve of WO$_{107}$ for V magnitudes reduced to the zero phase angle with parameter $G$ = 0.13 for observations at the Skalnat\'{e} Pleso on December 2 and 8. The lightcurves for the R band are shifted to the V lightcurves.}
\label{fig:ligcurve5}
\end{figure}

On December 2 and 8 the asteroid was observed in BVR filters with the 61-cm telescope at the Skalnat\'{e} Pleso.
The obtained color indices are: $B$--$V$ = $0.83 \pm 0.06$ mag; $V$--$R$ = $0.36 \pm 0.04$ mag for December 2; and $B$--$V$ = $0.84 \pm 0.06$ mag, $V$--$R$ = $0.37 \pm 0.04$ mag for December 8.
These values are in marginal agreement with the BART (FRAM-ORM) values.
The composite lightcurves on December 2 and 8 for reduced magnitudes of the asteroid in the V and R filters were used to calculate the phase curve parameter $G$ = $0.13 \pm 0.03$ and the absolute magnitudes $H$(max) = $18.84 \pm 0.03$ mag and $H_\mathrm{R}$(max) = $18.48 \pm 0.02$ mag for the maximum of the lightcurve (Figure \ref{fig:ligcurve5}). 
The rotation period was determined to be $5.025 \pm 0.003$ hours. The period is in good agreement with the values obtained in \citep{franco2021} and \citep{warner2021}.
The maximal amplitude of the lightcurve at the phase angle 5.0 deg is $1.0 \pm 0.05$ mag.
The absolute magnitudes of WO$_{107}$ averaged over the full rotation cycle (the mean V-lightcurve value) are $H$ = $19.332 \pm 0.033$ mag.
The obtained absolute magnitude agrees with the value $H$ = $19.30$ listed on the MPC site. 

\section{Shape modeling}
\label{sec:shape}

Our photometric observations and the Goldstone radar data \citep{benner} can be used to derive the shape model of the asteroid. As the quantity of the data is limited, fine resolution of the shape model is unattainable, and we constrain ourselves to a simple bilobal ellipsoidal shape model.

\subsection{Modeling photometric data}
\label{sec:shape-photometric}

\begin{figure}
\centering
\begin{tikzpicture}
\node[] at (0,0) {\includegraphics[width=.3\textwidth]{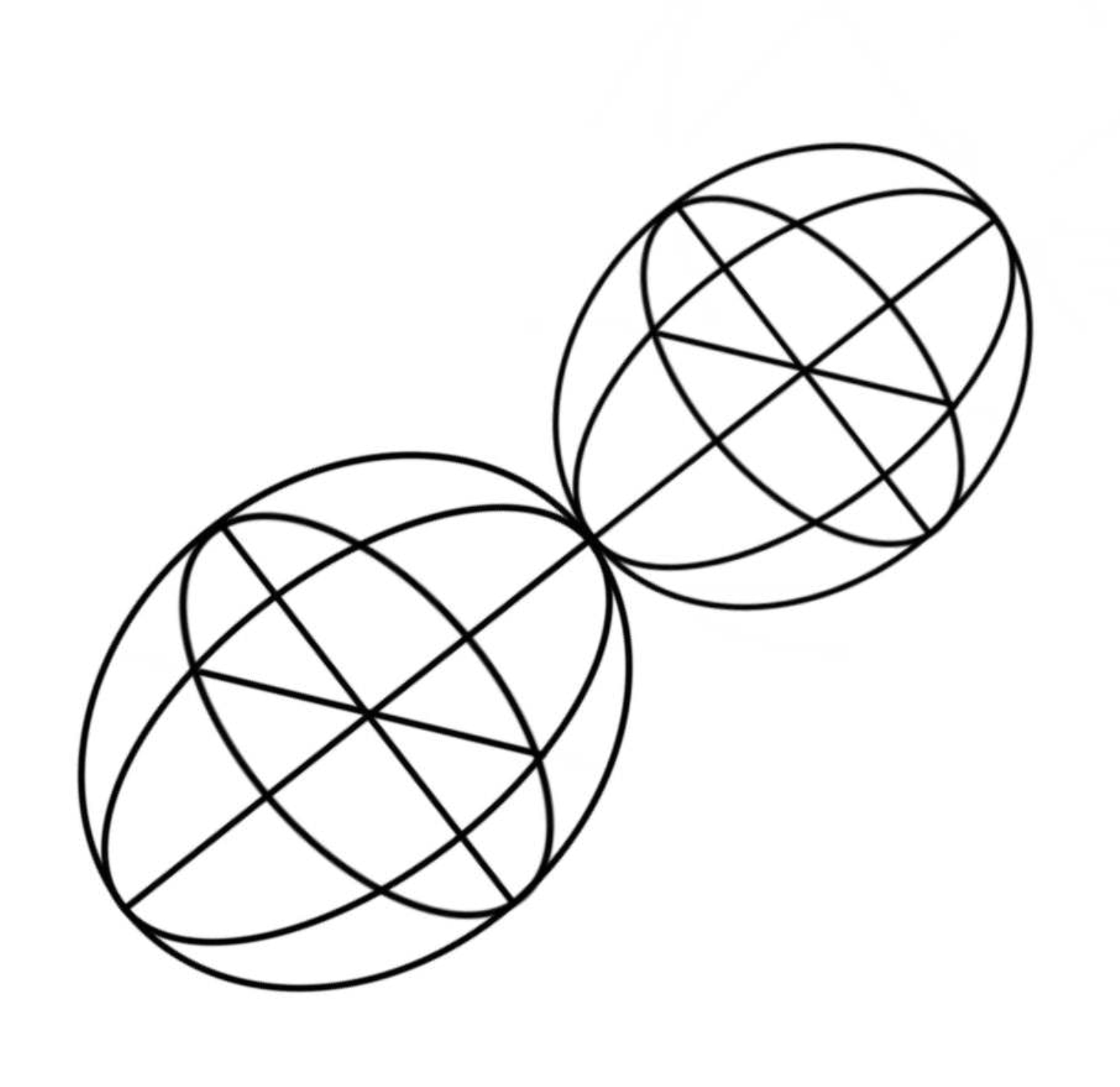}};
\node[red] at (-0.1,-0.4) {$a_1$};
\node[red] at (-1.4,-0.55) {$b_1$};
\node[red] at (-0.8,-1.25) {$c_1$};
\node[black!45!green] at (0.65,0.2) {$a_2$};
\node[black!45!green] at (0.85,1.1) {$b_2$};
\node[black!45!green] at (1.3,0.5) {$c_2$};
\end{tikzpicture}
\caption{Sketch for the shape model of WO$_{107}$ used in our simulations. It is composed of two ellipsoidal lobes.}
\label{fig:model}
\end{figure}
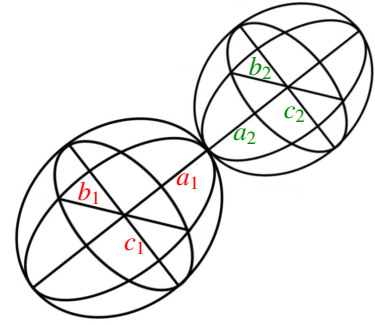

To model the rotation curve of the asteroid, we created a program that computes the brightness of an asteroid of a given shape and orientation. The details of this program are described in Appendix \ref{appendix:numerics}.

The asteroid is assumed to be composed of two ellipsoidal lobes with the semi-axes $a_1>b_1>c_1$ and $a_2>b_2>c_2$, whose longest semi-axes $a_1$ and $a_2$ are aligned with each other, while $c_1$ and $c_2$ are parallel to the rotation axis of the asteroid (Figure \ref{fig:model}). The asteroid rotates with the period $P$ around the rotation axis that points at the ecliptic coordinates $(\lambda,\beta)$. The rotation phase at the initial epoch is $\phi_0$.

Our program computes the brightness of the asteroid for each point of each lightcurve. Then we compute the residual between the theoretical and the observed lightcurves:
\begin{equation}
\chi^2_\mathrm{phot}=\frac{1}{2}\sum\limits_{i=1}^{N^\mathrm{lc}}\sum\limits_{j=1}^{N^\mathrm{p}_i}\frac{(m_{ij}^\mathrm{th}-m_{ij}^\mathrm{obs}-s_i)^2}{\Delta m_{ij}^2}.
\end{equation}
Here, $m_{ij}^\mathrm{obs}$ is the observed apparent magnitude of the asteroid, $\Delta m_{ij}$ is the observational error, and $m_{ij}^\mathrm{th}$ is the theoretical apparent magnitude, determined from the light flux using Pogson's formula. The index $i$ lists all the $N^\mathrm{lc}$ observed lightcurves, and the index $j$ lists all the $N^\mathrm{p}$ points in each lightcurve. All the observations are treated as relative photometry, and each lightcurve is shifted vertically by a value $s_i$, determined in such a way as to minimize $\chi^2$, i.e., from the condition $\partial\chi^2/\partial s_i=0$.
The null-point of Pogson's formula during the computation of $m^\mathrm{th}$ is taken arbitrarily, as its influence is canceled by the shifts $s_i$.

The free parameters of the model $a_1$, $b_1$, $c_1$, $a_2$, $b_2$, $c_2$, $\lambda$, $\beta$ and $\phi_0$ are updated by employing a Markov Chain Monte Carlo (MCMC) algorithm to minimize $\chi^2_\mathrm{phot}$. For this purpose, we use the $\tt emcee$ package \citep{emcee}, which implements the ensemble sampler. In our case, the ensemble sampler consists of three movers: StretchMove, DEMove and DESnookerMove, which generate proposals for updating the coordinates of the walkers in the ensemble. The mover is chosen randomly for each step, with probabilities of 0.4, 0.3, and 0.3, respectively. The mixture of moves is chosen for an efficient sampler suited for a high-dimensional model. The sampler used 120 walkers with 50000 steps to explore the parameters space, of which the first 2000 steps were discarded as a burn-in.

\subsection{Modeling radar data}

Delay--Doppler imaging of WO$_{107}$ was conducted from Goldstone on November 28, 29 and 30, 2020 \citep{benner}. The images published on the website represent in arbitrary units the distribution of the reflected power received by the radar as a function of time delay $\Delta t$ and frequency shift $\Delta\nu$.

To derive the parameters of the asteroid from its radar images, we created a program to numerically model the delay--Doppler data for a given asteroid shape. The asteroid is again represented by two connected triaxial ellipsoids (see Figure \ref{fig:model}), and a number of rays are cast on the asteroid and traced back. The numeric routine is similar to the one described in Appendix \ref{appendix:numerics}. One of the differences is that the direction of the incoming and outgoing rays are now both aligned with the asteroid--Earth direction. Another difference is that prior to the summation of the energies of all the rays we perform their binning by $\Delta t$ and $\Delta \nu$. The third difference is another scattering law, which for radar is chosen to be proportional to $\cos^n\theta$, with $\theta$ being the incidence angle, and $n$ being a free parameter \citep{virkki,magri}. Based on the results of preliminary optimization runs, the best results in terms of fitting data were achieved for $1.75\leq n\leq2.25$, which is close to the Lambert law. Thus, a value for $n$ was set to 2. Then the program determines the best values of free parameters of the asteroid model, and as a result, we obtain theoretical delay--Doppler images.

This simulated radar image is then compared to the observed one. For such comparison, the two images are shifted to align their centers of mass (the median energy-weighted $\Delta t$ and $\Delta \nu$). The corresponding vertical and horizontal shifts are both done by a rounded integer number of pixels, but this cannot cause a substantial error as the pixel size is much finer than the size of the asteroid. As we only had access to the published radar images, we tried to transform the modeled data correspondingly to facilitate their comparison. As the published images had saturated pixels, we achieved the same effect by setting a maximal pixel brightness, adjusted to reach the best visual similarity between the modeled and observed data. Then the shifted and saturated theoretical image is compared to the observed one pixel by pixel, and $\chi^2_\mathrm{rad}$ is computed as the sum of squared differences between the brightnesses over all the pixels of all the images.

The resulting inconsistency of radar images $\chi^2_\mathrm{rad}$ is used in an MCMC routine alongside the previously defined inconsistency of photometric lightcurves $\chi^2_\mathrm{phot}$. The ultimate MCMC fit to the data is performed via minimizing the sum $\chi^2=k\chi^2_\mathrm{rad}+\chi^2_\mathrm{phot}$, where the coefficient $k$ is introduced to equalize the contribution from photometric and radar data. It is empirically chosen in such a way that the residual values of $k\chi^2_\mathrm{rad}$ and $\chi^2_\mathrm{phot}$ for the best-fit models are nearly the same.

\subsection{Resulting shape}
\begin{figure}
\centering
\includegraphics[width=0.5\linewidth]{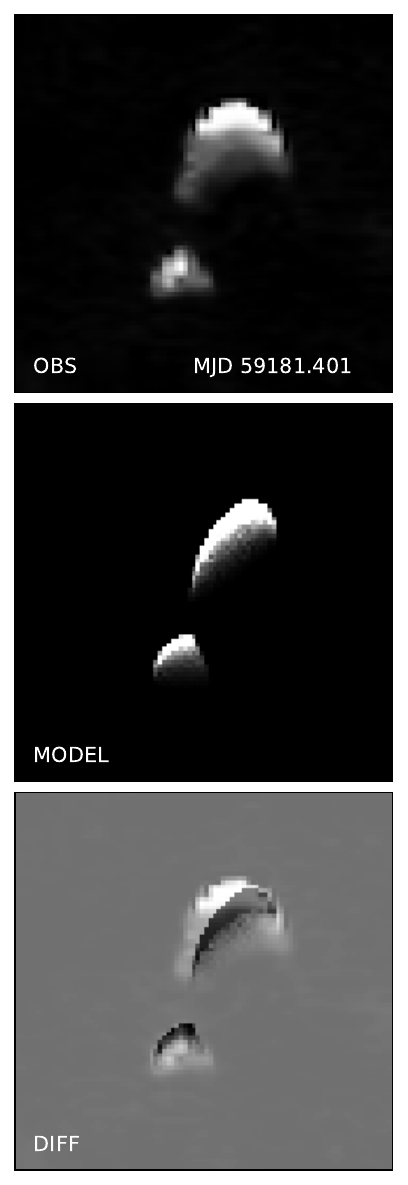}
\caption{A sample radar image. The first image represents the observed data, the second shows the modeled data, and the third one illustrates the difference between them.  
In case of a perfect agreement between the images, the pixels of the third image would have been gray, but in the case of worse agreement, they split into separate shapes.}
\label{fig:radar-red-green}
\end{figure}

As a result of the MCMC simulation, we obtain the distribution of the best-fit parameters of the asteroid. From all the values tried by the MCMC algorithm, we find the one that has the lowest $\chi^2$, and then take it for the starting value to further improve $\chi^2$ using the optimization method `differential evolution' based on \cite{Storn1997} from the package \textit{scipy.optimize} \footnote{SciPy package: https://docs.scipy.org/}.

A sample comparison between the theoretical model and one of the observed radar images is shown in Figure \ref{fig:radar-red-green}, and a comparison with the remaining images is shown in Figure \ref{fig:appendix-radar} in Appendix \ref{appendix:MCMC}. The theoretical image is computed for the asteroid model that has the lowest found $\chi^2$.

Comparison between the observed and simulated lightcurves of the asteroid is illustrated in Figure \ref{fig:appendix-lc} in Appendix \ref{appendix:MCMC}. Once again, the simulations are conducted for the asteroid model with the lowest $\chi^2$. The H magnitude shown in the lightcurve is the relative magnitude.

We see that the best-fit model describes well both the photometric and radar observational data. The photometric lightcurves are well-fitted in terms of their phase, amplitude, and overall shape. The minima of the lightcurves produced by self-shadowing and self-occultations of the asteroid are the most sensitive to the asteroid's shape and orientation, thus, they are the most crucial indicator of the accuracy of the fit. We see that in most of the cases, the minima of the lightcurves are satisfactorily reproduced by the best-fit theoretical model. The comparison of the observed and modeled radar images demonstrates a general agreement in shape, size and orientation, although each picture in the figures has disagreeing parts as an unavoidable result of the limitations of our theoretical model and imperfections of the observational data.

\begin{table}[]
\centering
\caption{Parameters of the asteroid}
\begin{tabular}{l|c|c|c}
Parameter & \multicolumn{2}{c|}{Value} \\
\hline
$a_1$ [km] & \multicolumn{2}{c|}{$0.34 \pm 0.04$} \\
$b_1$ [km] & \multicolumn{2}{c|}{$0.19 \pm 0.03$} \\
$c_1$ [km] & \multicolumn{2}{c|}{$0.18 \pm 0.05$} \\
$a_2$ [km] & \multicolumn{2}{c|}{$0.22 \pm 0.03$} \\
$b_2$ [km] & \multicolumn{2}{c|}{$0.21 \pm 0.03$} \\
$c_2$ [km] & \multicolumn{2}{c|}{$0.08 \pm 0.05$} \\
$P$ [h] & \multicolumn{2}{c|}{$5.017 \pm 0.002$} \\
& Solution 1 & Solution 2 \\
$\lambda$ [deg] & $96 \pm 8$ & $286 \pm 11$\\
$\beta$ [deg] & $-78 \pm 1$ & $-76 \pm 2$ \\
$\phi_0$ [deg] & $112 \pm 22$ & $301 \pm 65$ \\
\end{tabular}
\label{tab:MCMC results}
\end{table}

\begin{figure*}
\centering
\includegraphics[width=0.99\linewidth]{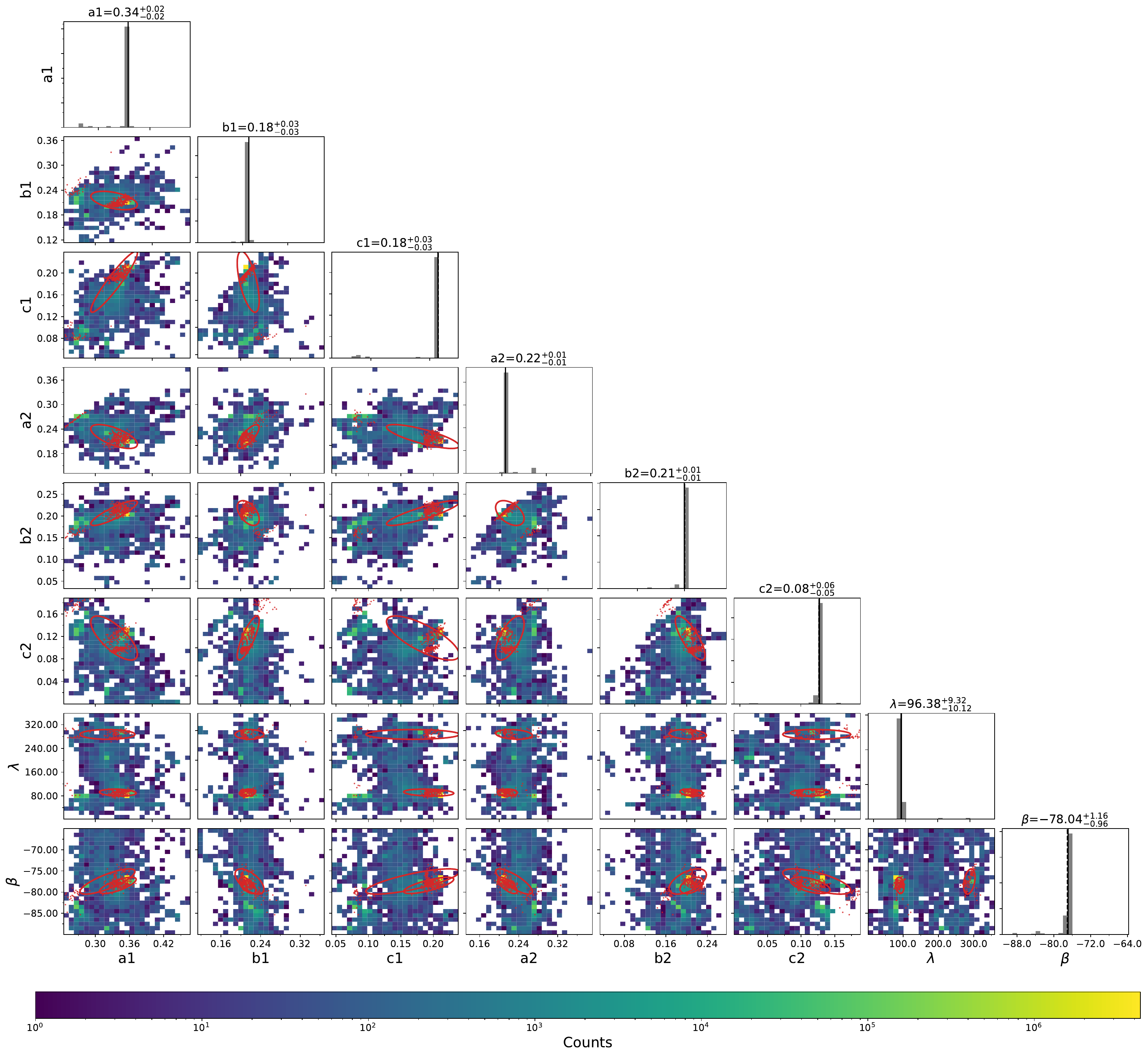}
\caption{Corner-plot showing the correlation between the most important parameters of the MCMC model of the asteroid. Red ellipses show the 1$\sigma$-contour, assuming a 2D normal distribution of errors, with the moments of this distribution estimated from the bootstrapped points. In the two bottom rows of panels there are two red ellipses corresponding to two different pole solutions.}
\label{fig:corellation}
\end{figure*}

The cornerplot in Figure \ref{fig:corellation} demonstrates the correlations between the results of the MCMC sampling of the six parameters that characterize the asteroid shape and the two angles that determine the orientation of its rotation axis. The diagonal subplots show the histograms of the distribution of walkers over each parameter, whereas the non-diagonal elements show the correlation in each pair of parameters.

\begin{figure}
\centering
\includegraphics[width=0.9\linewidth]{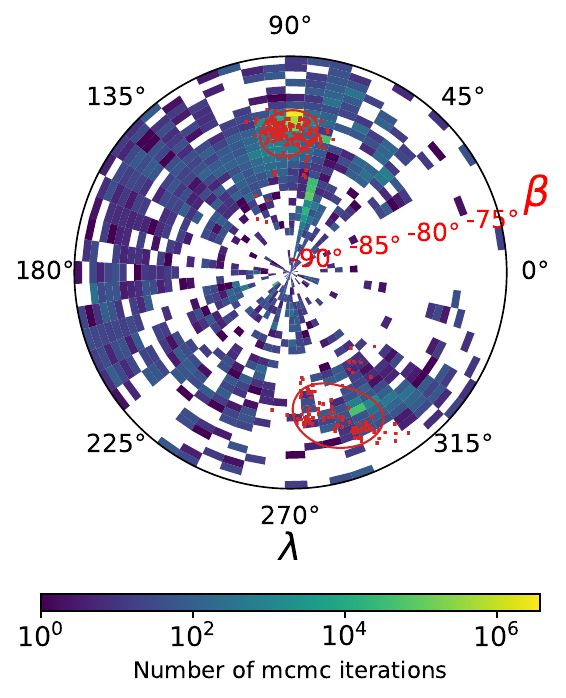}
\caption{Solutions for the pole orientation of the asteroid. The colormap shows the distribution of MCMC samples, while the red dots are bootstrap solutions.}
\label{fig:pole-solution}
\end{figure}

The distribution of walkers in the MCMC simulation (Figure \ref{fig:corellation}) is sensitive to the fine details of the MCMC algorithm and in general does not sample well the error distribution of the parameters. To get proper error estimates, we harness the bootstrapping method. For this sake, we reject a randomly selected half of the lightcurves and a randomly selected half of the radar images, and repeat the MCMC procedure to find the least-$\chi^2$ model of this bootstrapped subsample. We repeat this bootstrapping procedure 300 times with statistically independent subsamples. The distribution of the best-fit parameters for the bootstrapped subsamples is shown in Figure \ref{fig:corellation}.

Figure \ref{fig:pole-solution} shows the corresponding pole solution in the vicinity of the south pole of the ecliptic coordinates. It is a polar representation of the $\lambda$--$\beta$ plot, whose Cartesian representation is shown as the 7th panel in the last row in Figure \ref{fig:corellation}. We see on Fig. \ref{fig:pole-solution} that both the distribution of the MCMC walkers (color coded) and the best solution for each bootstrap (red points) cluster around two separate pole solutions, one centered around $\lambda=96^\circ$, $\beta=-78^\circ$, the other one around $\lambda=286^\circ$, $\beta=-76^\circ$. Such non-unique pole solutions often occur in the shape modeling of asteroids \citep{Kaasalainen2006}. We split the solutions into two groups, $0^\circ<\lambda<180^\circ$ and $180^\circ<\lambda<360^\circ$ respectively, which visually approximately correspond to the two clusters of the bootstrapped points. Then, assuming a 2D normal distribution of errors, we compute the 1$\sigma$-contour in each of the two groups of points, and mark the contours with red ellipses. The solution at $\lambda=96^\circ$, $\beta=-78^\circ$ (reached in 65\% out of 300 bootstrapped subsamples) is more plausible than the solution at $\lambda=286^\circ$, $\beta=-76^\circ$ (35\% of subsamples), but neither of the solutions can be ruled out.

Table \ref{tab:MCMC results} shows the best-fit parameters of the MCMC simulation and their errors. The errors are computed as the mean-squared deviation of the best-fit parameters for the bootstrapped subsamples with respect to the full data sample.

\section{Density determination}
\label{sec:density}
The consideration of the gravitational equilibrium of two triaxial ellipsoids in contact allows us to determine the density of the asteroid.

The density estimate is obtained from the condition that the gravitational attraction between the two lobes of the asteroid is equal to their centrifugal repulsion. The gravitational attraction between the two lobes of equal densities is
\begin{equation}
F_\mathrm{gr}=\frac{G}{(a_1+a_2)^2}\cdot\frac{4}{3}\pi\rho a_1 b_1 c_1 \cdot\frac{4}{3}\pi\rho a_2 b_2 c_2.
\end{equation}
Here, we treat the gravitational field of both ellipsoidal lobes as the gravitational field of a point mass. The accuracy of this approach is sufficient for our purposes, as we will see that the resulting density error is much greater than the error that one could expect to stem from this assumption.

The distance from the second lobe to the center of mass of the binary system is
\begin{equation}
r_2=(a_1+a_2)\frac{a_1 b_1 c_1}{a_1 b_1 c_1+a_2 b_2 c_2}.
\end{equation}
The centrifugal force acting on this lobe is
\begin{equation}
F_\mathrm{centr}=\omega^2 m_2 r_2=\left(\frac{2\pi}{P}\right)^2 \frac{4}{3}\pi\rho a_2 b_2 c_2 \frac{(a_1+a_2)a_1 b_1 c_1}{a_1 b_1 c_1+a_2 b_2 c_2},
\end{equation}
where $\omega$ is the angular velocity of the asteroid and $m_2$ is the mass of the second lobe.

We assume the force balance, $F_\mathrm{centr}=F_\mathrm{gr}$. In reality, this condition can break in both directions. First, the centrifugal force can slightly exceed the gravitational force, $F_\mathrm{centr}>F_\mathrm{gr}$, due to the tensile strength of the asteroid material in the contact region of the two lobes. Second, the gravity can exceed the centrifugal force,
$F_\mathrm{centr}<F_\mathrm{gr}$, due to the contact pressure between the two lobes. Still, the clear contact binary shape seen on the radar images implies a small area of contact between the two lobes and thus at most a small difference between the gravity and the centrifugal force. Equating the two, one gets the density of the asteroid, 
\begin{equation}
\rho=\frac{3\pi(a_1+a_2)^3}{GP^2(a_1 b_1 c_1+a_2 b_2 c_2)}.
\label{density}
\end{equation}
If all the dimensions of the asteroid are multiplied by the same factor $k$, then both numerator and denominator in Eq. (\ref{density}) get multiplied by $k^3$, thus $\rho$ remain unchanged. In this sense, the resulting asteroid density does not depend on the absolute values of the ellipsoid axes, just on their ratios.

\begin{figure}
\centering
\includegraphics[width=0.99\linewidth]{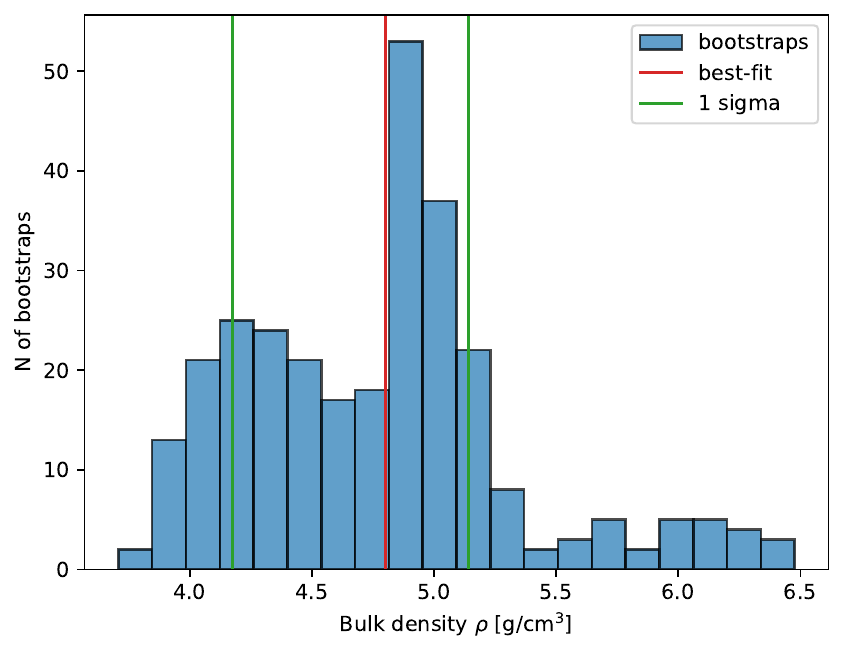}
\caption{Density estimates for WO$_{107}$. The blue histogram shows the distribution of densities calculated from bootstraps. The red line shows the density estimate from the best-fit asteroid model for the full data, and the green ones show $1\sigma$ uncertainty range estimated from the bootstrapped data.}
\label{fig:density}
\end{figure}

We apply this equation to the asteroid shapes and rotation periods obtained for each of 300 bootstrapped solutions, and show the results in Figure \ref{fig:density}. The estimations for $1 \sigma $ density error are shown with vertical green lines, while the density for the lowest $\chi^2$ value for the full sample is shown with the thick vertical line, and we assume it for the nominal value. The plot hints at a bimodal distribution, but we cannot suggest any simple cause for this apparent bimodality. The resulting density estimated from the shape is $\rho=4.80^{+0.34}_{-0.63}$ g/cm$^3$.

\section{Origin and orbital evolution}

To get information about the possible origin of the asteroid we used the NEOPOP software, which is based on the debiased distribution model of near-Earth asteroids \citep{granvik2018}. The model suggests $\nu_{6}$ secular resonance with Saturn to be the only possible source of the asteroid. In fact, $\nu_{6}$ is one of the major contributors to the near-Earth asteroid population, responsible for nearly 25$\%$ bodies on highly eccentric orbits similar to WO$_{107}$ in magnitude \citep{nesvorny2023}.

To obtain information about the orbital evolution of WO$_{107}$, we performed numerical simulations. As the base for computation, we used GENGA -- a hybrid symplectic N-body integrator, which utilizes graphical processing units to integrate planet and small body dynamics \citep{genga2014, genga2022}. 

The base of the model consisted of all the planets of the Solar System and the most massive bodies of the main asteroid belt: Ceres, Vesta, and Pallas. The asteroid was represented by 5,000 and 200,000 clones, which were integrated $10^6$ years to the past and $10^4$ years to the near future, respectively. The Yarkovsky effect was included in the simulation.

\begin{figure}
\centering
\includegraphics[width=0.49\textwidth]{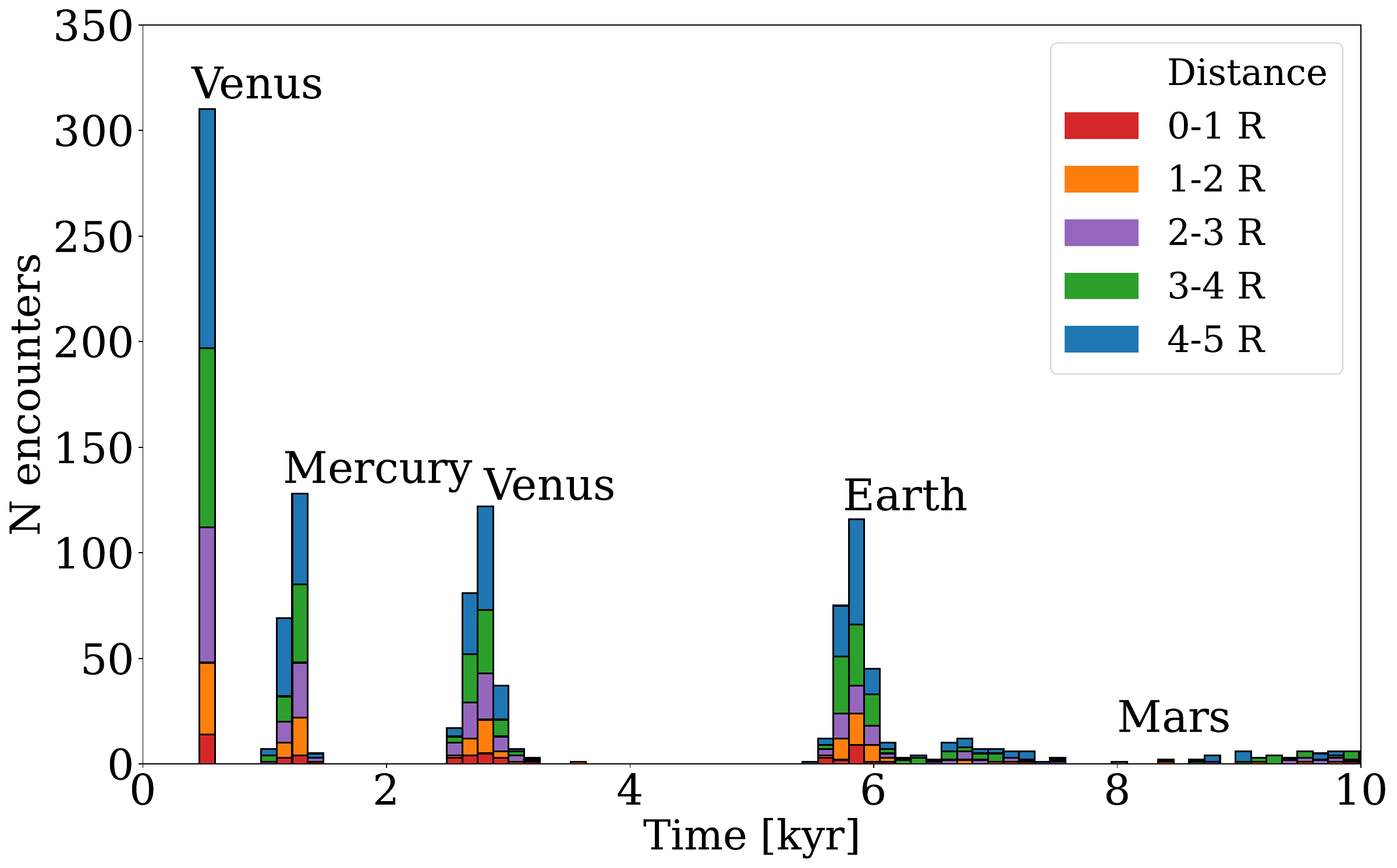}
\caption{Encounters of WO$_{107}$ with terrestrial planets in the simulation to 10 kyr into the future.}
\label{fig:encounters}
\end{figure}

The results of the modeling show that WO$_{107}$ experiences numerous close encounters with the terrestrial planets throughout its orbital history as a consequence of its highly eccentric orbit. For the nearest 200 years into the future, integrations indicate minor interactions with Venus and Earth, which slightly raise the orbit, followed by close encounters with Venus at 500 years mark, Mercury at 1,200, Venus at 2,700, Earth at 5,600-6,000, and Mars at 7,000-10,000 years (Figure \ref{fig:encounters}). In terms of risk assessment of WO$_{107}$ as a potentially hazardous asteroid, only 250 out of 200,000 clones (0.1 $\%$) have close encounter distances of $\leq 5 R_\mathrm{Earth}$, and 14 clones (0.007 $\%$) collide with the Earth (distances of $\leq R_\mathrm{Earth}$). 

\label{sec:origin}

\section{Discussion and conclusions}
\label{sec:results}
We carried out observations of WO$_{107}$ from November 28 to December 6, 2020, which showed lightcurves typical for contact binary asteroids. The precise rotation period and maximum amplitudes of brightness variations at a wide range of phase angles from 5 up to 68 degrees were determined.

The asteroid lightcurves and its radar data strongly suggested a contact binary shape. Therefore, we assumed that the asteroid is composed of two ellipsoidal lobes, fitted their parameters to the observational data, and obtained a satisfactory agreement. It served as a justification of the chosen shape model, although more complicated asymmetric shape models cannot be discredited. The best-fit shape model consists of the ellipsoidal lobes with the sizes $0.68\times 0.38 \times 0.36$ km and $0.44 \times 0.42 \times 0.16$ km. The asteroid rotates around its shortest axis with a rotation period of $5.017\pm 0.002$ hr. The most probable solution for the angular velocity vector of the asteroid points at the ecliptic coordinates $\lambda=96^\circ \pm 8^\circ$ and $\beta=-78^\circ \pm 1^\circ$, whereas another less probable solution around $\lambda=286^\circ \pm 11^\circ$, $\beta=-76^\circ \pm 2^\circ$ cannot be disregarded. Such uncertainty of $\lambda$ is largely explained by the value of $\beta$ that is close to $-90^\circ$, so that the ratio of dimensions of the uncertainty area on the celestial sphere is much less extreme, as it can be seen in Fig. \ref{fig:pole-solution}, which gives the pole-on view of the possible positions of the asteroid pole on the celestial sphere.

From the consideration of the gravitational equilibrium of two triaxial ellipsoids in contact, we estimated the density for the asteroid, $\rho=4.80^{+0.34}_{-0.63}$ g/cm$^3$. Interestingly, the flattening of the asteroid lobes has a strong influence on this high density: considering two spheres the result would have been about 3 times smaller. The resulting mass of the asteroid can be estimated as $2.96^{+0.46}_{-0.14}\cdot 10^{11}$ kg. The obtained high density of the asteroid can imply a substantial metal content. 
This agrees with the possible M taxonomic type of the asteroid.

The orbital simulation of a large number of clones of WO$_{107}$ shows that it is a potentially hazardous asteroid. Numerical simulations for the next 10,000 years give the integral probability of colliding with the Earth of $7\cdot 10^{-5}$, with the maximum around 5,900 years into the future.

\begin{acknowledgements}
This work was partially funded by the National Research Foundation of Ukraine, grant No. 2020.02/0371 ``Metallic asteroids: search for parent bodies of iron meteorites, sources of extraterrestrial resources''. YK and OG thank for supporting the European Federation of Academies of Sciences and Humanities (grants ALLEA EFDS-FL1-18 and ALLEA EFDS-FL1-16, respectively) and Astronomical Observatory Institute of Faculty of Physics of Adam Mickiewicz University, where a part of the work was done. OI, MH were supported by the Slovak Grant Agency APVV no. APVV-19-0072, the Slovak Grant Agency VEGA 2/0059/22. VL acknowledges financial support from the German Excellence Strategy via the Heidelberg Cluster of Excellence (EXC 2181 - 390900948) ``STRUCTURES''. The authors are grateful to A. Zhornichenko, V. Agletdinov, A. Novichonok for the help with observations and S. Mykhailova for assistance with the data reduction. The operation of FRAM-ORM telescope is supported by grants of the Ministry of Education of the Czech Republic LM2023032 and LM2023047, as well as EU/MEYS grants CZ.02.1.01/0.0/0.0/16\_013/0001403, CZ.02.1.01/0.0/0.0/18\_046/0016007,  CZ.02.1.01/0.0/0.0/ 16\_019/0000754,
and CZ.02.01.01/00/22\_008/0004632. Based on data from the CMC15 Data Access Service at CAB (INTA-CSIC). Survey (APASS), funded by the Robert Martin Ayers Sciences Fund. This work has made use of data from the Asteroid Terrestrial-impact Last Alert System (ATLAS) project. ATLAS is primarily funded to search for near-Earth asteroids through NASA grants NN12AR55G, 80NSSC18K0284, and 80NSSC18K1575; byproducts of the NEO search include images and catalogs from the survey area. The ATLAS science products have been made possible through the contributions of the University of Hawaii Institute for Astronomy, the Queen's University Belfast, the Space Telescope Science Institute, and the South African Astronomical Observatory. YK thanks the French PAUSE program, which
provides support to scientists at risk.  
We are very grateful to the anonymous Referee, whose insightful critical remarks helped very much to improve the contents and the style of the article.  The authors express their gratitude to all those people who defend 
Ukraine and thus made it possible to prepare this article. 
\end{acknowledgements}

\appendix

\section{Computation of the asteroid brightness}
\label{appendix:numerics}

We created a program that separates the binary asteroid into multiple small facets, checks which facets are illuminated by the Sun and are seen from the Earth, and adds up the scattered intensity of all the facets to calculate the total brightness of the asteroid.

For each point on the lightcurve, we use the JPL Horizons database to obtain the vectors connecting the asteroid with the Sun and the Earth. 
These vectors are then transformed from the ecliptic coordinate frame to the body-fixed coordinate frame with the equations inverse to Eq. (1) in \cite{damit}.

In the body-fixed frame, the asteroid shape model is composed of two ellipsoids, aligned with the coordinate axes and shifted along the $x$ axis, so that the radius vector of a point on one of the ellipsoids is given in parametric form by
\begin{equation}
\mathbf{r}=(x,y,z)=(d+a\cos\beta\cos\alpha,b\cos\beta\sin\alpha,c\sin\beta).
\end{equation}
Here, $a$, $b$, and $c$ are the semimajor axes of the ellipsoid, $d$ is the shift of the ellipsoid from the center of mass of the asteroid, $\alpha\in[0;2\pi)$ and $\beta\in[-\pi/2;\pi/2]$ are two variables that parametrize the ellipsoid surface.
To cover the entire surface, we sample $\alpha$ and $\beta$ with small steps $\delta\alpha=\delta\beta=\frac{\pi}{2N}$, where $N$ is the number of points per $\pi/2$ radians, which in our simulations was set $N = 30 - 50$. We compute the linear parts of the changes of the radius vector between the points adjacent in longitudinal and latitudinal directions,
\begin{equation}
\mathbf{\delta r_\alpha}=\frac{\partial \mathbf{r}}{\partial \alpha}\delta\alpha=(-a\cos\beta\sin\alpha,b\cos\beta\cos\alpha,0)\delta\alpha,
\end{equation}
\begin{equation}
\mathbf{\delta r_\beta}=\frac{\partial \mathbf{r}}{\partial \beta}\delta\beta=(-a\sin\beta\cos\alpha,-b\sin\beta\sin\alpha,c\cos\beta)\delta\beta.
\end{equation}
The surface element $\mathbf{\delta S}=\mathbf{\delta r_\alpha}\times \mathbf{\delta r_\beta}$ based on vectors $\mathbf{\delta r_\alpha}$ and $\mathbf{\delta r_\beta}$ is given by
\begin{equation}
\mathbf{\delta S}=(bc\cos^2\beta\cos\alpha,ac\cos^2\beta\sin\alpha,ab\cos\beta\sin\beta)\delta\alpha\,\delta\beta.
\end{equation}
From $\mathbf{\delta S}$ we determine the surface area $\delta S=|\mathbf{\delta S}|$ and the normal vector $\mathbf{n}=\mathbf{\delta S}/\delta S$ of each surface element.

Then the brightness created by the surface element is computed as a combination of the Lommel--Seeliger and Lambert
laws in accordance with the first equation in Section 2.3 of \cite{damit}. The Lambertian part in this equation is assumed equal to $0.1$, as it is done for most of the asteroids in the DAMIT database\footnote{Astronomical Institute of the Charles University, Josef Ďurech, Vojtěch Sidorin, DAMIT, https://astro.troja.mff.cuni.cz/projects/damit}. As we only use relative photometry, we ignore the phase function, which gives just a vertical shift to a lightcurve but no noticeable change in its shape. The angles in the scattering laws are computed based on the normal vector $\mathbf{n}$ of the surface element and the previously computed asteroid--Earth and asteroid--Sun vectors in the body-fixed frame.

The brightness created by each surface element is added to the total brightness of the asteroid only if the element is simultaneously illuminated by the Sun and seen from the Earth. It means that both the incoming and the outgoing ray lie above the local horizon and do not intersect the other lobe of the asteroid. The requirement that they lie above the local horizon is taken care of by the scattering law, whereas the condition of not intersecting the other lobe must be checked separately. To check whether a ray intersects an ellipsoid, we perform an affine transformation that turns the ellipsoid into a sphere, and then check the condition of whether the transformed ray intersects the sphere.

Thus, our program accounts for shadowing but not self-illumination of the asteroid. Still, the latter is not expected to be too important, as the albedo of the asteroid is low.

The program was subject to multiple unit tests to check if it works correctly. For one big spherical lobe and the second lobe of negligibly small size the brightness remained constant. The same was true if one spherical lobe was situated fully inside the other spherical lobe (a nonphysical case used purely as a mathematical test). For two equal spherical lobes at very large separation (wide binary asteroid), the brightness was twice as big as for one lobe. At zero phase angle the computed brightness for a Lambertian sphere agreed with the analytic result. For an arbitrary contact binary asteroid at 0 phase angle the brightness was 0. For a contact binary with the lobes 2:1:1 and 1:1:1 in equatorial aspect the lightcurve agreed with our physical expectations. In the polar aspect at 0 phase angle the lightcurve was constant, the same as the maximum in the equatorial aspect at 0 phase angle. It was also checked that the simultaneous revolution of the asteroid axis and the vectors asteroid--Sun and asteroid--Earth do not change the lightcurve. These unit tests made us sufficiently confident that the program works correctly.

\section{MCMC results}
\label{appendix:MCMC}
In this Appendix, we present the extended plots from the optimization simulations of the asteroid parameters, samples of which were demonstrated in the main text of the article.

\begin{figure*}
\centering
\includegraphics[width=1.0\textwidth]{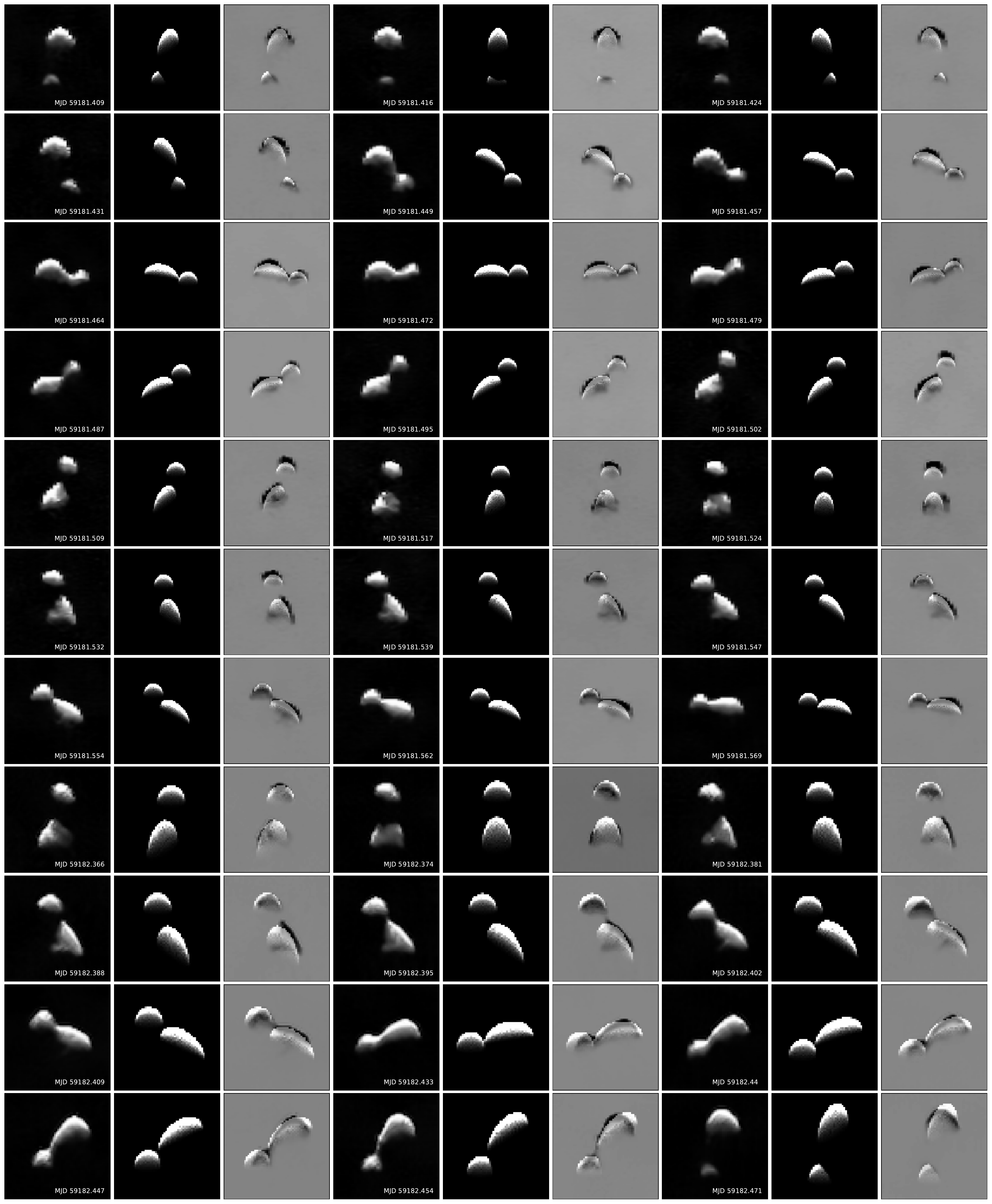}
\caption{Observed and simulated radar images and their difference. Observed images are taken from \cite{benner}.}
\label{fig:appendix-radar}
\end{figure*}

In Figure \ref{fig:appendix-radar}, we show the observed images in which the observation time is indicated as modified Julian date (MJD), simulated radar images from our numeric model, and the difference between observed and modeled images. They are visualized in the same manner as in Figure \ref{fig:radar-red-green} (see explanations of it in the main text). The total number of radar images used in our analysis was 71: 1 in Figure \ref{fig:radar-red-green} and 70 in Figure \ref{fig:appendix-radar}.

\begin{figure*}
\centering
\includegraphics[width=1.0\textwidth]{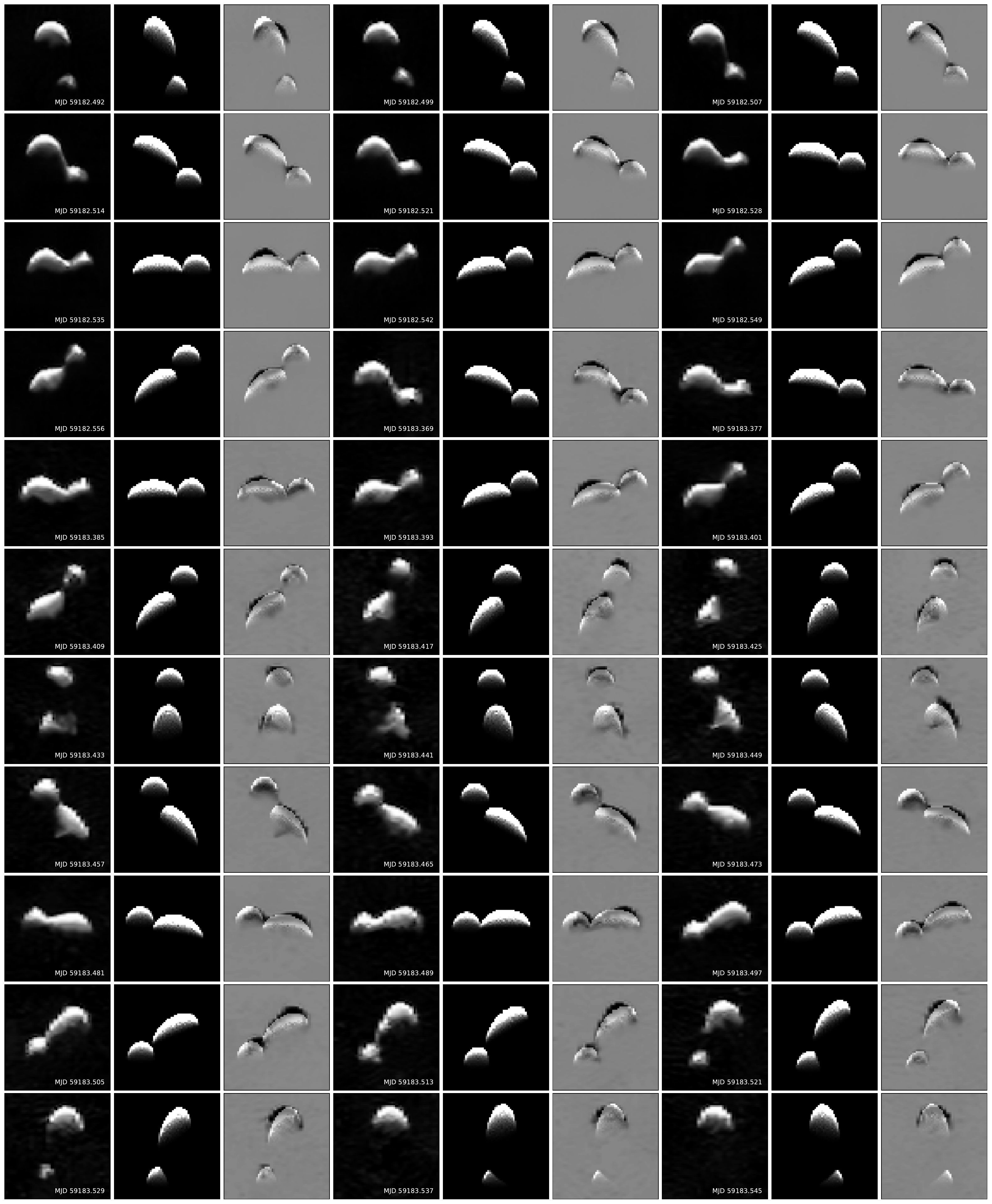}
\caption{Observed and simulated radar images and their difference. Continuation of Fig. \ref{fig:appendix-radar}}
\label{fig:appendix-radar2}
\end{figure*}

\begin{figure*}
\centering
\includegraphics[width=0.93\textwidth]{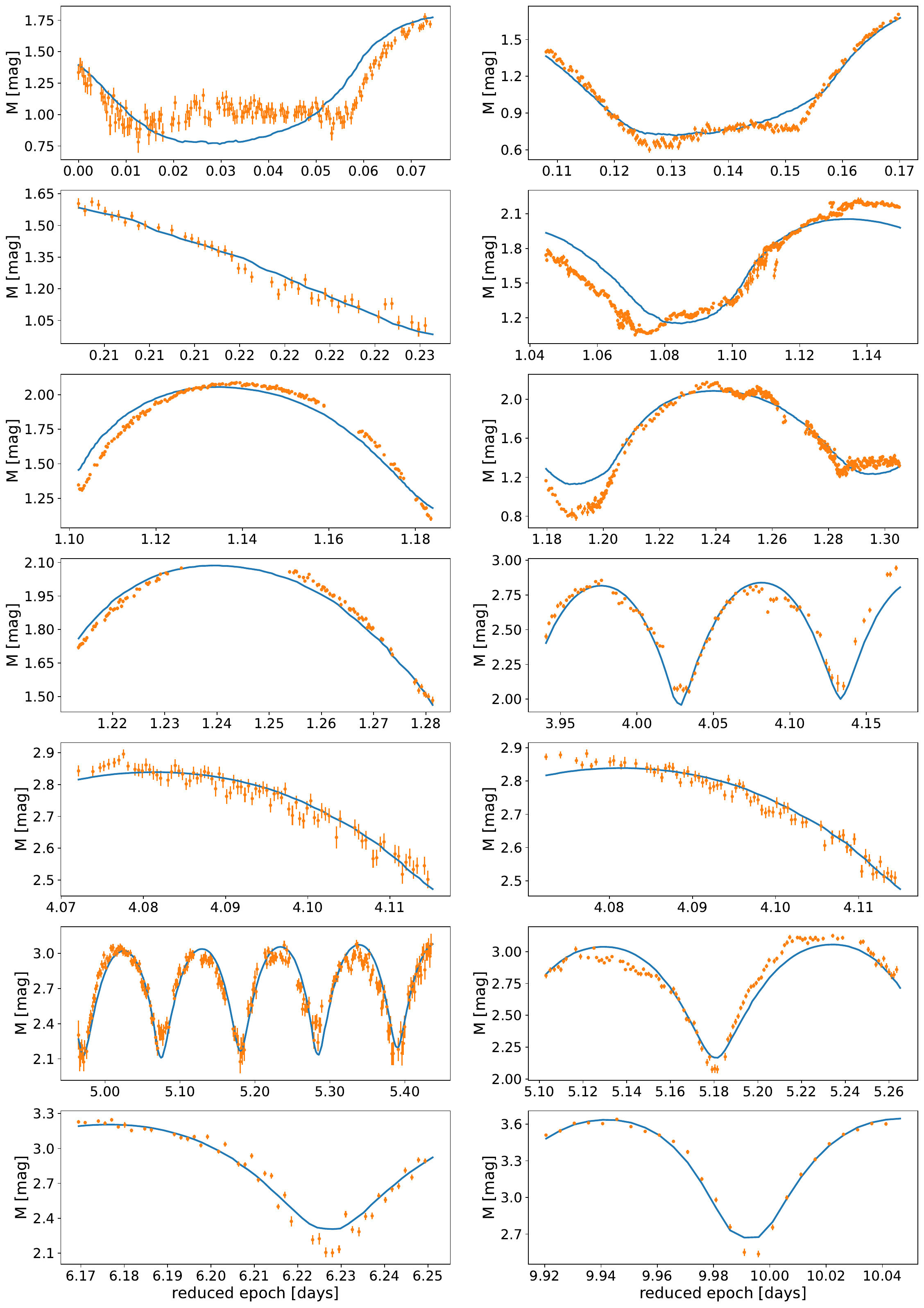}
\caption{Comparison of observed (orange) and simulated (blue) photometric data. Zero date is JD 2459182.3299.}
\label{fig:appendix-lc}
\end{figure*}

Figure \ref{fig:appendix-lc} shows all the observed lightcurves for the asteroid overlaid with modeled ones.

\end{document}